\documentclass[11pt]{article}
\usepackage{moriond,epsfig}

\def\be{\begin{equation}}
\def\ee{\end{equation}}
\def\bea{\begin{eqnarray}}
\def\eea{\end{eqnarray}}

\begin{document}
\vspace*{3.0cm}
\title{Recent results on anomalous $J/\psi$ suppression in Pb-Pb collisions at 158 AGeV/c}
\author{Helena Santos for the NA50 Collaboration\\
B.~Alessandro$^{11}$, C.~Alexa$^{4}$, R.~Arnaldi$^{11}$, M.~Atayan$^{13}$, C.~Baglin$^{2}$, 
A.~Baldit$^{3}$, M.~Bedjidian$^{12}$, S.~Beol\`e$^{11}$, V.~Boldea$^{4}$, P.~Bordalo$^{7,a}$, 
S.R.~Borenstein$^{10,b}$, G.~Borges$^{7}$, A.~Bussi\`ere$^{2}$, L.~Capelli$^{12}$, C.~Castanier$^{3}$, 
J.~Castor$^{3}$, B.~Chaurand$^{10}$, B.~Cheynis$^{12}$, E.~Chiavassa$^{11}$, C.~Cical\`o$^{5}$, 
T.~Claudino$^{7}$, M.P.~Comets$^{9}$, S.~Constantinescu$^{4}$, P.~Cortese$^{1}$, J.~Cruz$^{7}$, 
A.~DeFalco$^{5}$, N.~DeMarco$^{11}$, G.~Dellacasa$^{1}$, A.~Devaux$^{3}$, S.~Dita$^{4}$, 
O.~Drapier$^{10}$, B.~Espagnon$^{3}$, J.~Fargeix$^{3}$, P.~Force$^{3}$, M.~Gallio$^{11}$, 
Y.K.~Gavrilov$^{8}$, C.~Gerschel$^{9}$, P.~Giubellino$^{11,c}$, M.B.~Golubeva$^{8}$, 
M.~Gonin$^{10}$, A.A.~Grigorian$^{13}$, S.~Grigorian$^{13}$, J.Y.~Grossiord$^{12}$, 
F.F.~Guber$^{8}$, A.~Guichard$^{12}$, H.~Gulkanyan$^{13}$, R.~Hakobyan$^{13}$, R.~Haroutunian$^{12}$, 
M.~Idzik$^{11,d}$, D.~Jouan$^{9}$, T.L.~Karavitcheva$^{8}$, L.~Kluberg$^{10}$, A.B.~Kurepin$^{8}$, 
Y.~Le Bornec$^{9}$, C.~Louren\c co$^{6}$, P.~Macciotta$^{5}$, M.~Mac~Cormick$^{9}$, 
A.~Marzari-Chiesa$^{11}$, M.~Masera$^{11}$, A.~Masoni$^{5}$, M.~Monteno$^{11}$, A.~Musso$^{11}$, 
P.~Petiau$^{10}$, A.~Piccotti$^{11}$, J.R.~Pizzi$^{12}$, W.L.~Prado da Silva$^{11,e}$, F.~Prino$^{1}$, 
G.~Puddu$^{5}$, C.~Quintans$^{7}$, L.~Ramello$^{1}$, S.~Ramos$^{7,a}$, P.~Rato Mendes$^{7}$, 
L.~Riccati$^{11}$, A.~Romana$^{10}$, H.~Santos$^{7}$, P.~Saturnini$^{3}$, E.~Scalas$^{1}$, 
E.~Scomparin$^{11}$, S.~Serci$^{5}$, R.~Shahoyan$^{7,f}$, F.~Sigaudo$^{11}$, M.~Sitta$^{1}$, 
P.~Sonderegger$^{6,a}$, X.~Tarrago$^{9}$, N.S.~Topilskaya$^{8}$, G.L.~Usai$^{5}$, E.~Vercellin$^{11}$, 
L.~Villatte$^{9}$, N.~Willis$^{9}$ and T.~Wu$^{9}$}
\address{
$^{~1}$ Universit\`a del Piemonte Orientale, Alessandria and INFN-Torino, Italy
$^{~2}$ LAPP, CNRS-IN2P3, Annecy-le-Vieux,  France
$^{~3}$ LPC, Univ. Blaise Pascal and CNRS-IN2P3, Aubi\`ere, France
$^{~4}$ IFA, Bucharest, Romania
$^{~5}$ Universit\`a di Cagliari/INFN, Cagliari, Italy
$^{~6}$ CERN, Geneva, Switzerland
$^{~7}$ LIP, Lisbon, Portugal
$^{~8}$ INR, Moscow, Russia
$^{~9}$ IPN, Univ. de Paris-Sud and CNRS-IN2P3, Orsay, France
$^{10}$ Laboratoire Leprince-Ringuet, Ecole Polytechnique and CNRS-IN2P3, Palaiseau, France
$^{11}$ Universit\`a di Torino/INFN, Torino, Italy
$^{12}$ IPN, Univ. Claude Bernard Lyon-I and CNRS-IN2P3, Villeurbanne, France\\ 
a) also at IST, Universidade T\'ecnica de Lisboa, Lisbon, Portugal, 
b) on leave of absence from York College, CUNY, New York, USA, 
c) also at CERN, Geneva, Switzerland, 
d) also at Faculty of Physics and Nuclear Techniques, University of Mining and Metallurgy, Cracow, Poland, 
e) now at UERJ, Rio de Janeiro, Brazil, f) on leave of absence from YerPhI, Yerevan, Armenia}
\maketitle\abstracts{
We report our latest results on charmonium suppression as measured at the CERN-SPS  in  Pb-Pb interactions at 
an incident beam momentum of 158~GeV/c per nucleon.
Preliminary results obtained from the most recent sample of data collected in year 2000 under improved experimental
conditions are compared with published results. For the most peripheral Pb-Pb collisions, $J/\psi$ suppression  
agrees with the normal absorption measured from interactions of lighter nuclei while a steady increasing abnormal
supression is observed with increasing centrality.}

\section{Introduction}

The study of charmonium production in ultrarelativistic heavy ion collisions 
by experiments NA38 and NA50 has been motivated by the
search 
for a phase transition of nuclear matter from its normal state to a deconfined quark gluon plasma,
as predicted to occur, under extreme energy densities or 
temperature conditions, by non-perturbative QCD. In such a deconfined matter, 
the $J/\psi$ is expected  to be suppressed by Debye colour screening \cite{MatSatz}. 
As the vector-meson is produced at the very early stages of the collision, its leptonic decay provides 
a probe of these stages, since this mode is insensitive to strong interactions and, therefore, 
to the hadronization stage.\\[-7mm]

\section{Data selection and analysis method}
The main element of the NA50 detector is a muon spectrometer. 
It is complemented by centrality detectors which measure the neutral transverse 
energy $E_T$, the beam spectators energy $E_{ZDC}$ and the charged multiplicity produced in 
the interaction. Several beam monitoring devices are also used for various purposes. 
A detailed description can be found in~\cite{NA50_B410_327}. The goal of 
the year 2000 Pb-Pb data collection is to investigate, in particular, the most peripheral collisions
which, in previous data samples, could be contaminated by Pb-air interactions. 
The data have been taken, therefore, with the target placed in the vacuum pipe.\\
The event selection is done according to the following criteria: dimuons are selected in the 
rapidity window $ 2.92 \leq y_{lab} < 3.92 ~(0 \leq y_{CM} < 1)$ and with a Collins-Soper angle
$\mid cos\theta_{CS}~\mid~<~0.5$, leading to an acceptance of around 14\% in the mass region of interest. 
Parasitic interactions of an incident Pb-ion in the Beam Hodoscope and upstream from the target are rejected 
by a BH interaction detector and by Anti-halo counters. On-target interactions are selected requiring a good 
correlation between hits in the two planes of the Multiplicity Detector. 
Double interactions are discarded by a shape analysis of the signal in the Electromagnetic Calorimeter. 
Residual pileup events are rejected by a 2$\sigma$ cut in the $E_T-E_{ZDC}$ correlation~\cite{Quintans}.
Off-target muons are excluded by a cut on the transverse distance between the extrapolated muon track 
and the beam line.\\
In each centrality bin of the analysis, the opposite-sign dimuon mass spectrum is adjusted 
in the pertinent mass region, accounting for the different ingredients: $J/\psi$, $\psi'$ and the continuum 
formed by Drell-Yan, $D\bar{D}$ and the combinatorial background from $\pi$ and $K$ decays. The shapes of 
the resonances are obtained by Monte Carlo simulation followed by reconstruction with the same program as
used for the real data analysis. Drell-Yan and $D\bar{D}$ are generated at leading order by Pythia and treated 
by the NA50 simulation-reconstruction chain. The combinatorial background is properly extracted from 
like-sign pairs through the formula $N_{BG}=2 \sqrt{N^{++}N^{--}}$. \\
\vspace*{-7mm}

\section{Before the year 2000 -- The state of the art}
\begin{figure}[!ht]
\begin{center}
 \begin{minipage}[h]{.38\linewidth}
  \includegraphics[width=\linewidth]{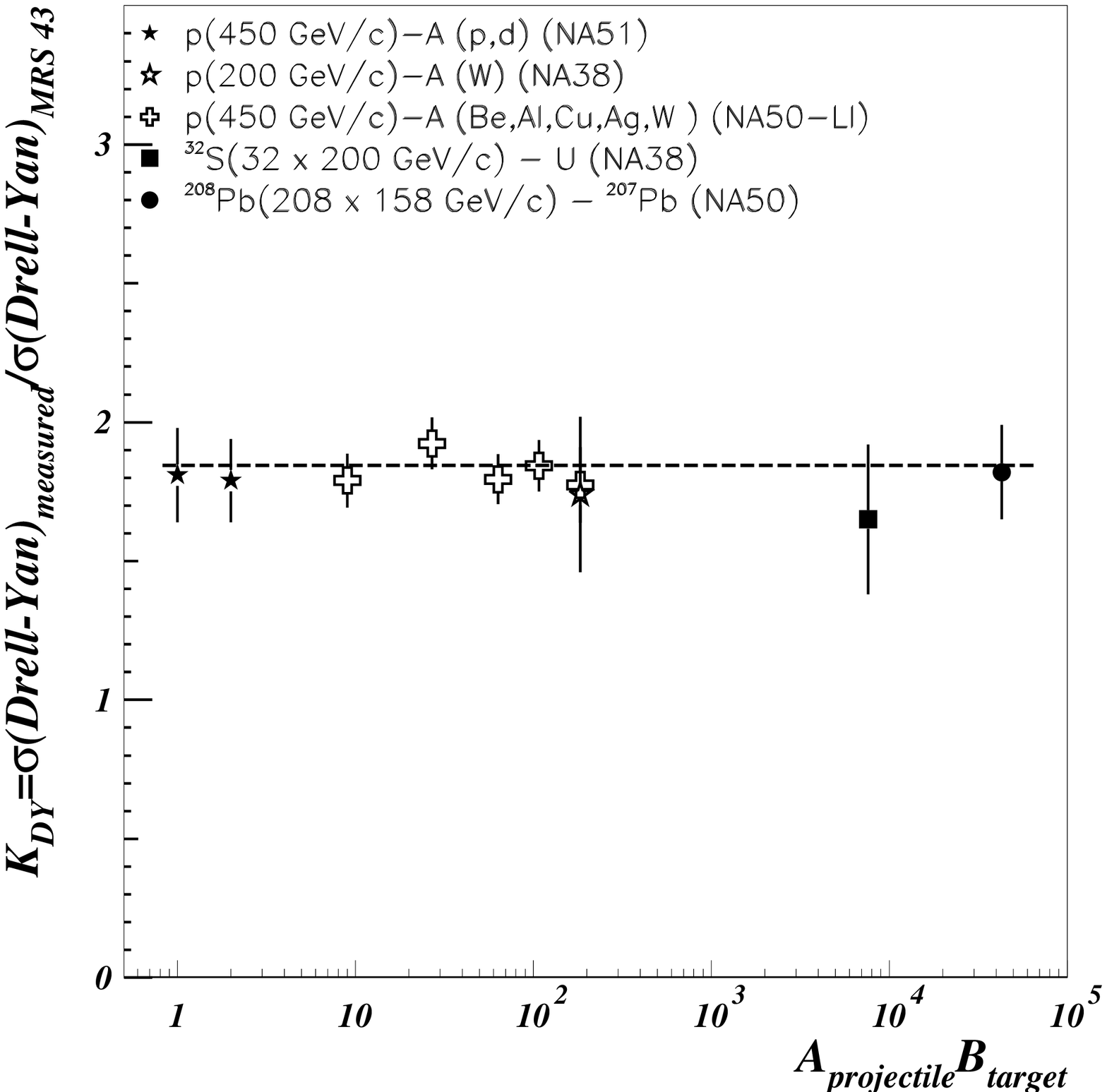}
  \vspace*{-9mm}
  \caption{Drell-Yan K-factor as a function of {$\rm A_{projectil}~\times~B_{target}$}}
  \label{fig:kdy}
 \end{minipage} 
\hspace*{6mm}
 \begin{minipage}[h]{.39\linewidth}
  \vspace*{-2mm}
  \includegraphics[width=\linewidth]{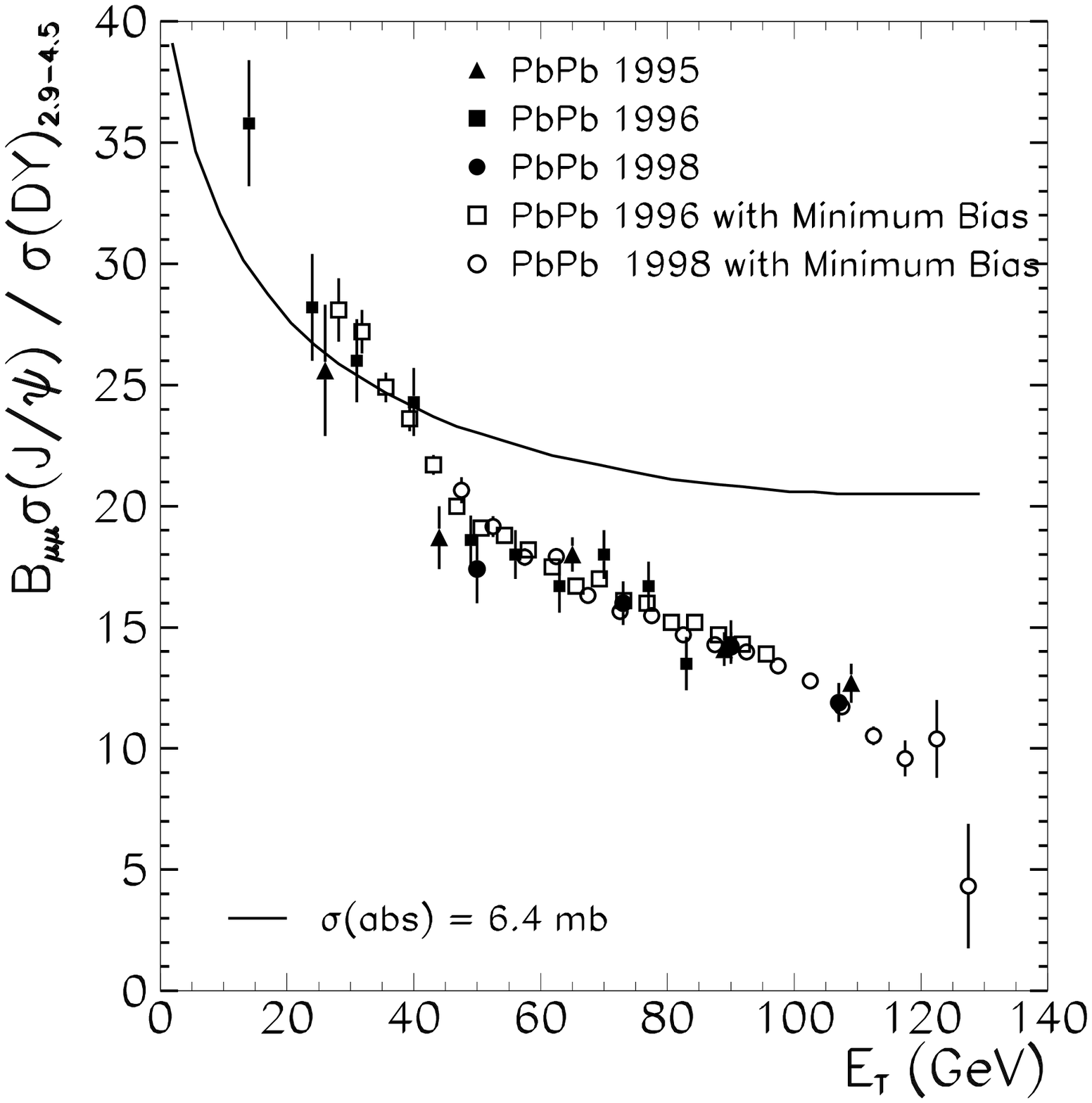}
  \vspace*{-10mm}
  \caption{$J/\psi/DY$ as a function of $E_T$; the absorption curve is a fit
  to the NA38 p-A and S-U data} 
  \label{fig:psidypubl}
\end{minipage} 
\end{center}
\end{figure}

The NA38/NA51/NA50 experiments have verified  that, as expected theoretically, 
the Drell-Yan cross-section $\sigma$(DY) is proportional to the number of elementary 
nucleon-nucleon collisions, from proton-proton up to lead-lead interactions included (see Fig.~\ref{fig:kdy}). 
The ratio of the cross-sections for $J/\psi$ and Drell-Yan production is therefore proportional
to the $J/\psi$ production cross-section per nucleon-nucleon collision. Moreover, 
the ratio $B_{\mu\mu}\sigma(J/\psi)/\sigma(DY)$ is directly measured in the so-called standard analysis method and
insensitive to reconstruction or trigger inefficiencies (identical muon pair topology for $J/\psi$ and 
Drell-Yan events) or to absolute normalization uncertainties which, obviously, cancel out.
Fig.~\ref{fig:psidypubl} shows the ratio $B_{\mu\mu}\sigma(J/\psi)/\sigma(DY)$ as a function of  $E_T$, 
for the previously collected
Pb-Pb data samples -- 1995~\cite{NA50_B410_327}$^,$\cite{NA50_B410_337}, 1996~\cite{NA50_B450_456} and 
1998~\cite{NA50_B477_28}. The closed points are the results obtained with the standard analysis method
described above while the open points are the 1996 and 1998 results obtained through the "Minimum Bias analysis"
method explained in [4,5]. The continuous line is the $J/\psi$ normal 
absorption in nuclear matter and fits the NA51 p-p and p-d and NA38 p-A and S-U results.  
A clear departure of the $B_{\mu\mu}\sigma(J/\psi)/\sigma(DY)$ ratios from this normal absorption curve is observed
at mid-centrality, 
suggesting the onset of an anomalous $J/\psi$ suppression mechanism. Futhermore, no saturation of the ratio 
at high $E_T$ is observed, as opposed to the absorption curve trend.
\vspace*{-4mm}

\vspace*{-12mm}

\section{The preliminary year 2000 results}
\begin{figure}[!ht]
\begin{center}
 \begin{minipage}[h]{.37\linewidth}
 \vspace*{-6mm}
  \includegraphics[width=\linewidth]{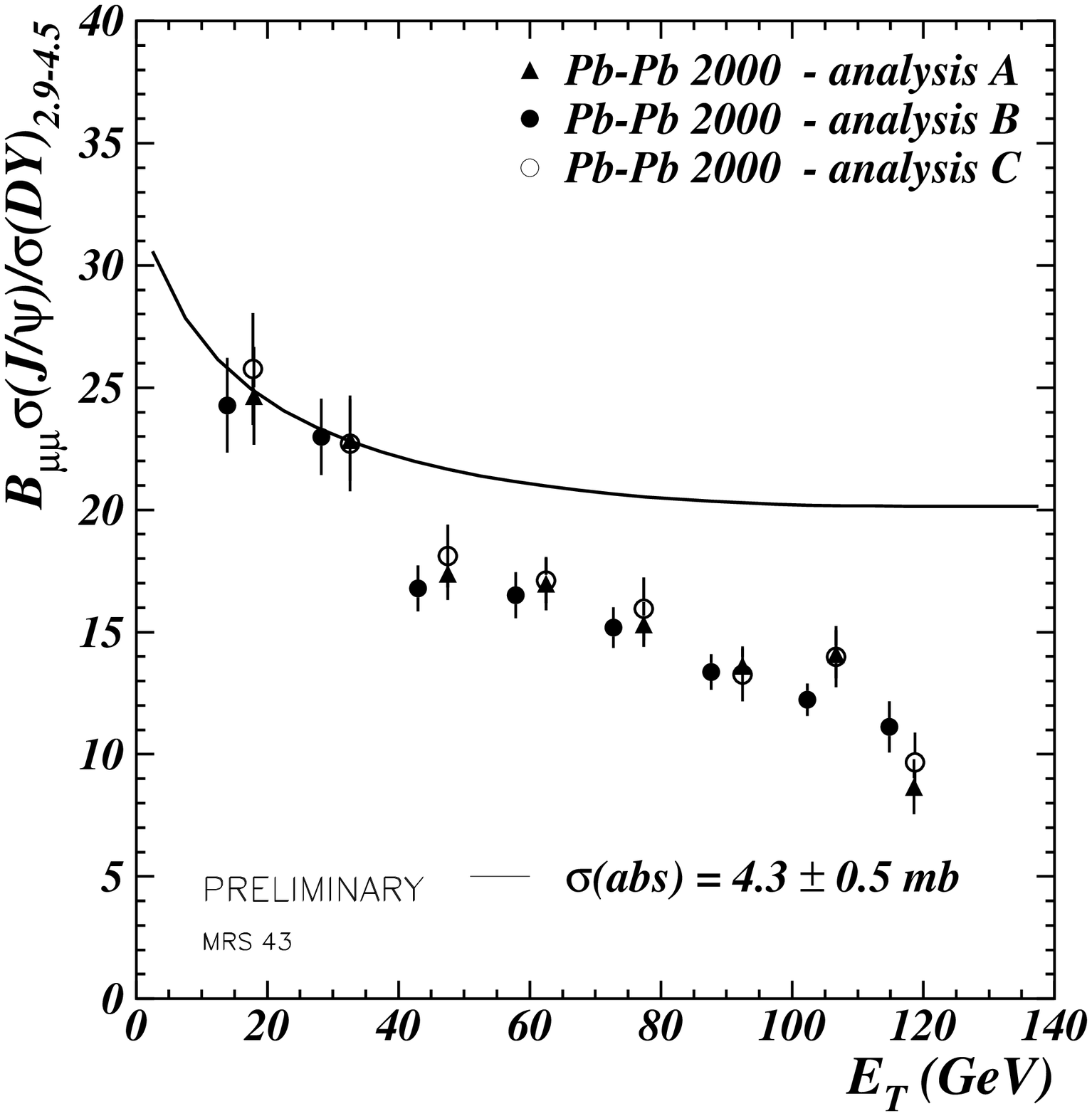}
  \vspace*{-11mm}
  \caption{$J/\psi/DY$ as a function of $E_T$}
  \label{fig:3mrs}
 \end{minipage}
 \hspace*{6mm}
 \begin{minipage}[h]{.38\linewidth}
  \vspace*{-8mm}
  \includegraphics[width=\linewidth]{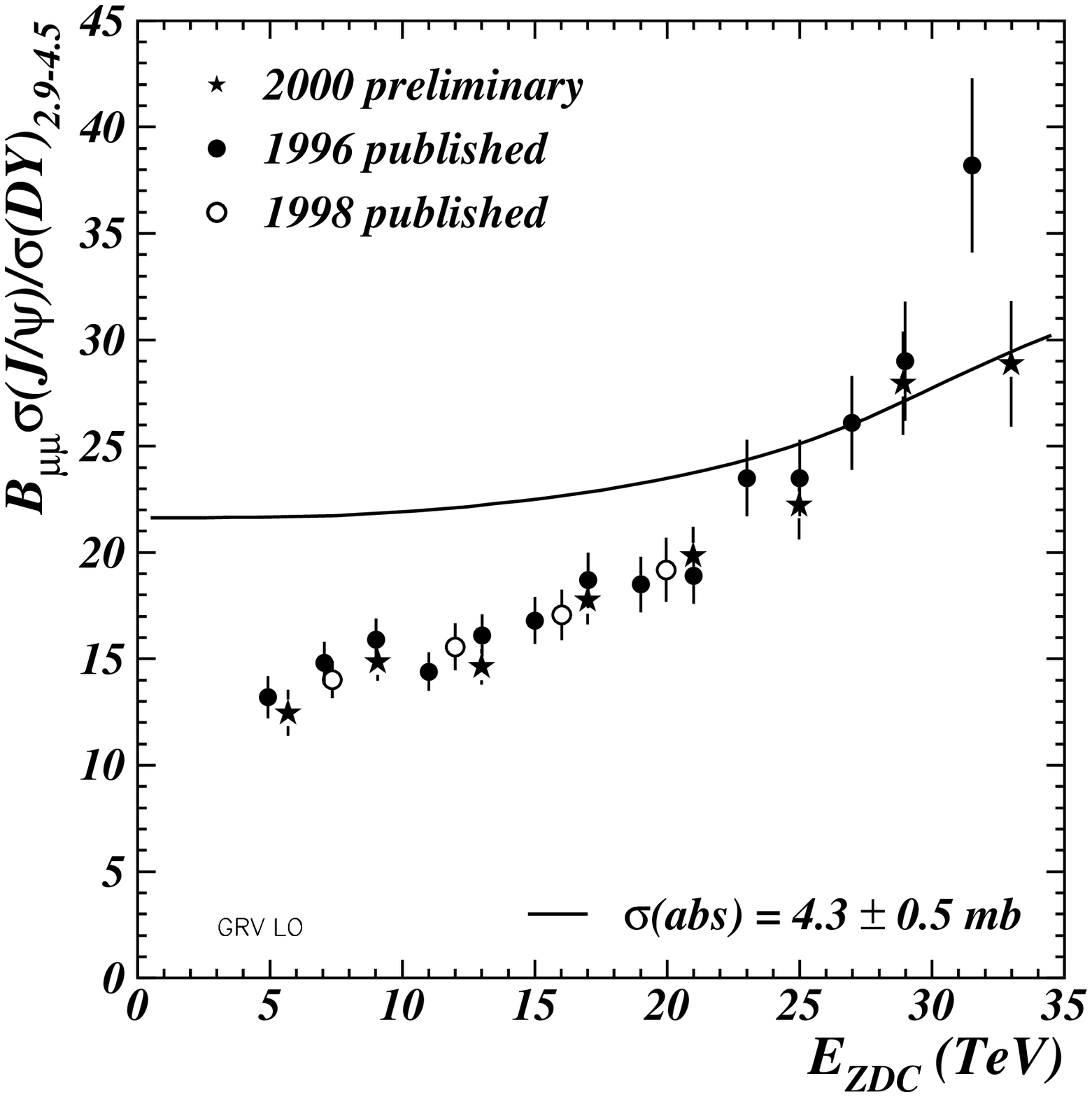}
  \vspace*{-11mm}
  \caption{$J/\psi/DY$ as a function of $E_{ZDC}$} 
  \label{fig:compzdcgrv}
\end{minipage} 
\end{center}
\end{figure}

\begin{figure}[!ht]
\begin{center}
 \begin{minipage}[h]{.37\linewidth}
  \vspace*{-4mm}
  \includegraphics[width=\linewidth]{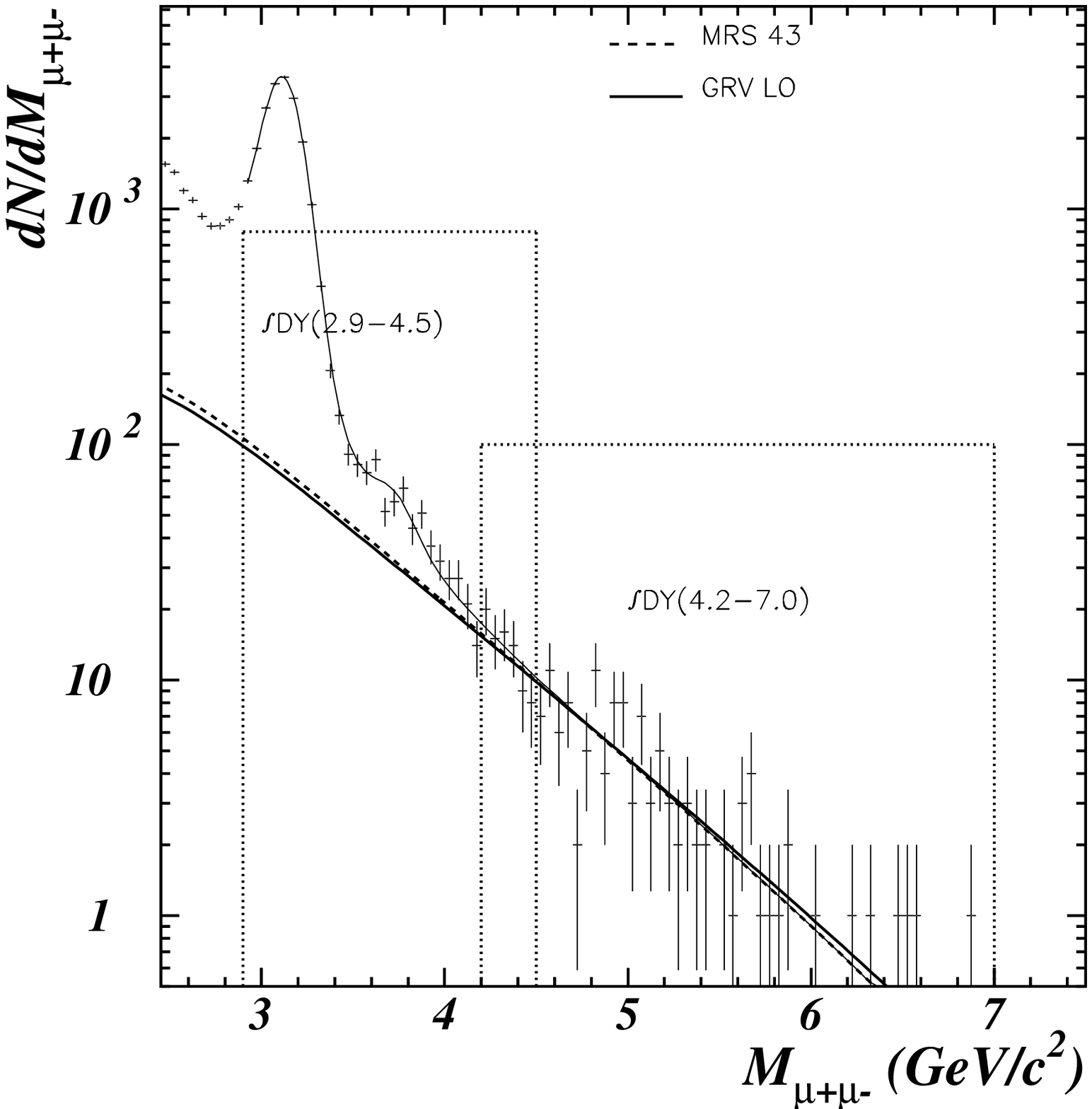}
  \vspace*{-10mm}
  \caption{The Drell-Yan mass distributions for MRS43 and GRV LO PDFs}
  \label{fig:stepcompare}
 \end{minipage}
 \hspace*{6mm}
 \begin{minipage}[h]{.37\linewidth}
  \vspace*{-4mm}
  \includegraphics[width=\linewidth]{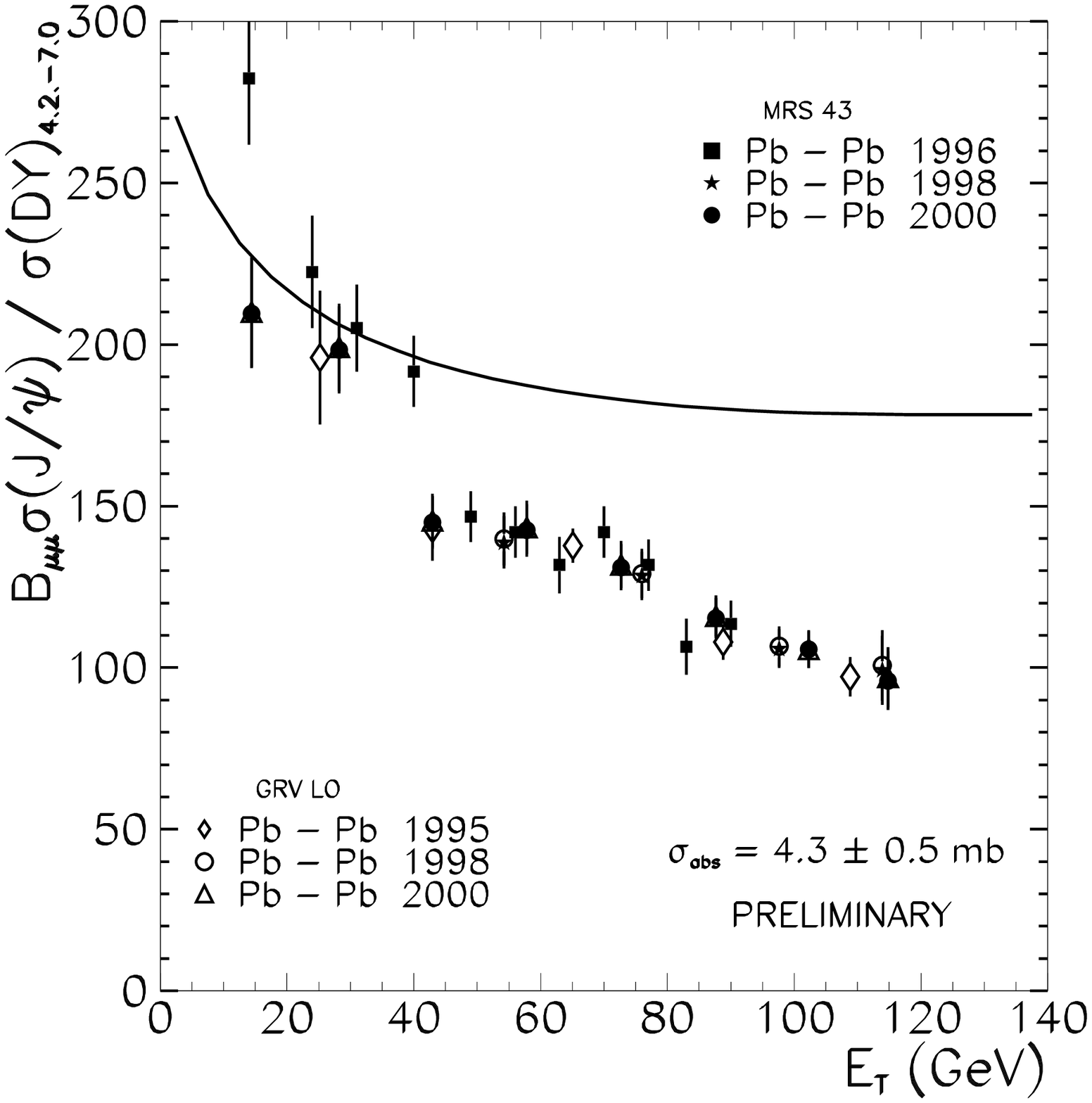}
  \vspace*{-10mm}
  \caption{$J/\psi/DY$ as a function of $E_T$} 
  \label{fig:tmundo}
\end{minipage} 
\end{center}
\end{figure}

The NA50 Collaboration has a new high statistics p-A data sample, obtained with a 450 GeV/c incident energy beam. 
Since the $J/\psi$ absorption cross sections for p-A and S-U collisions are compatible within the statistical errors, 
a simultaneous Glauber fit is performed which leads to the common value of 
$\sigma(J/\psi)_{abs}=4.3 \pm 0.5$ mb~~\cite{NA50_B553_167}. 
This result, combined with  the NA38 S-U data   reanalyzed in order to have the same analysis conditions, 
leads to an updated   
normal nuclear absorption. \\
From the above described fit to the dimuon invariant mass spectra, 
we extract the $J/\psi$ and the Drell-Yan yields (the latter in the mass interval 2.9 - 4.5 GeV/c$^2$) 
and obtain the ratio $B_{\mu\mu}\sigma(J/\psi)/\sigma(DY)$  and its dependence
as a function of $E_T$ or $E_{ZDC}$. 
Fig.~\ref{fig:3mrs} shows the agreement among 
independent analyses with slightly different data selections and fit methods. It also shows 
that peripheral collisions indeed follow the absorption curve. 
On the other hand, the Pb-Pb 2000 data confirm the departure from this normal behaviour of the 
$J/\psi/DY$ ratios at mid-centrality, already observed in the previously collected data. 
The results also confirm the steady decrease of the $J/\psi$ production rate at high $E_T$. 
The same $J/\psi$ suppression pattern is seen in 
Fig.~\ref{fig:compzdcgrv} when $E_{ZDC}$ is used as the centrality estimator. For this variable, the agreement
is quite good between new and already published results, 
except for the most peripheral collisions, 
probably due to Pb-air interactions in the 1996 setup conditions.\\
In the 2.9-4.5 GeV/c$^2$ invariant mass range, muons from $J/\psi$ and Drell-Yan have similar features, 
namely acceptance and momentum. Nevertheless, different PDFs chosen to generate Drell-Yan lead to different 
mass distributions, as shown in Fig.~\ref{fig:stepcompare}. The difference under the $J/\psi$ peak between 
the Drell-Yan cross sections generated with MRS43 and GRV LO amounts to 10\% systematic effect
in the ~$J/\psi/DY$ normalizations, independent of centrality. To overcome this uncertainty, 
the 4.2-7.0 GeV/c$^2$ mass range is chosen 
as the new $J/\psi$ suppression reference. In fact, 
beyond 4.2 GeV/c$^2$, the Drell-Yan pairs are the only contribution to the dimuon mass spectrum and thus 
both structure functions give the same normalization after adjustment to the data. 

Fig.~\ref{fig:tmundo} shows all NA50 results 
using the new mass range for Drell-Yan as reference. All data sets agree, regardless on the fact that Pb-Pb 1995 results were obtained using GRV LO PDFs while 
1996, 1998 and 2000 used MRS43. Pb-Pb 1998 and 2000 analyses with GRV LO is also shown.\\
\vspace*{-6mm}
\section{Conclusions}
In the most recently collected Pb-Pb data sample, the NA50 Collaboration has measured 
$B_{\mu\mu}\sigma(J/\psi)/\sigma(DY)$ as a function of  centrality .
With respect to previously collected data samples, the use of a setup with the target region under vacuum allows 
to extend, confidently, the analysis to very peripheral collisions.\\
These new data,  confirming previous analyses results, show that $J/\psi$ production follows the normal 
nuclear absorption pattern for peripheral Pb-Pb collisions, while exhibiting a clear departure for central 
collisions and suggesting, therefore, the onset of another  $J/\psi$ suppression mechanism.

\section*{Acknowledgments}
\vspace*{-1mm}
This work was partially supported by Funda\c{c}\~ao para a Ci\^encia e a Tecnologia.

\end{document}